\documentclass[a4paper]{jpconf}
\usepackage{graphicx}
\begin{document}
\title{Higgsless electroweak symmetry breaking at the LHC}

\author{Veysi Erkcan \"Ozcan$^1$}

\address{On behalf of the ATLAS and CMS Collaborations\\[1.5ex]
$^1$Department of Physics and Astronomy,
University College London, Gower Street, London WC1E~6BT, UK}

\ead{eo@hep.ucl.ac.uk}

\begin{abstract}
While the Higgs model is the best studied scenario of electroweak symmetry
breaking, a number strongly-coupled models exist, predicting new
signatures.  Recent studies of WW and WZ final states at the 
ATLAS and CMS experiments are summarized and expected sensitivities are
presented within the frameworks of the technicolor straw-man model
and the electroweak chiral Lagrangian.
\end{abstract}

\def\xspace     {\,}
\def\to         {\ensuremath{\rightarrow}\xspace}
\def\etal       {{\it et~al}\xspace}
\def\BR         {{\ensuremath{\cal BR}}}
\def\Lh		{{\ensuremath{\cal L}}}
\def\pt		{\ensuremath{p_{T}}}

\def\Zz      {\ensuremath{Z^0}}
\def\Z	     {\ensuremath{Z}\xspace}
\def\W	     {\ensuremath{W}}
\def\Wp      {\ensuremath{W^+}\xspace}
\def\Wpm     {\ensuremath{W^\pm}}
\def\rhotc   {\ensuremath{\rho_{TC}\xspace}}
\def\pitc    {\ensuremath{\pi_{TC}\xspace}}

\def\zbb      {\ensuremath{Zb\bar{b}}}
\def\tt       {\ensuremath{t\bar{t}}}
\def\afour    {\ensuremath{\alpha_{4}}}
\def\afive    {\ensuremath{\alpha_{5}}}

\def\fb    {\ensuremath{\mathrm{fb}}}
\def\invfb {\ensuremath{\mbox{\,fb}^{-1}}\xspace}
\newcommand{\tev}{\ensuremath{\mathrm{\,Te\kern -0.1em V}}\xspace}
\newcommand{\tevns}{\ensuremath{\mathrm{\,Te\kern -0.1em V}}}
\newcommand{\gev}{\ensuremath{\mathrm{\,Ge\kern -0.1em V}}\xspace}
\newcommand{\gevns}{\ensuremath{\mathrm{\,Ge\kern -0.1em V}}}

\def\npb   #1 #2 {{\it Nucl.~Phys.}~B~{\bf#1} #2}
\def\cpc   #1 #2 {{\it Comput.~Phys.~Commun.}~{\bf#1} #2}
\def\prp   #1 #2 {{\it Phys.~Rept.}~{\bf#1} #2}
\def\prd   #1 #2 {{\it Phys.~Rev.}~D~{\bf#1} #2}
\def\prl   #1 #2 {{\it Phys.~Rev.~Lett.}~{\bf#1} #2}
\def\plt   #1 #2 {{\it Phys.~Lett.}~{\bf#1} #2}
\def\pre   #1 {{\it Preprint}~#1}
\vspace{-0.37pc}

\section{Introduction}

Understanding the dynamics of electroweak symmetry breaking (EWSB)
will be a principal goal for the ATLAS and CMS experiments at the Large
Hadron Collider (LHC). While the physics of the EWSB in the minimal
Standard Model (perhaps extended by super-symmetry) is based on a purely
weakly-interacting Higgs sector~\cite{Higgs}, there is no fundamental reason
for the existence of a unique elementary scalar particle. The mechanism for
the symmetry breaking might rely on strong dynamics.

Unitarity violation in the scattering of longitudinal massive gauge bosons
sets the scale for such strong dynamics to
$\Lambda\sim 4\pi M_{W}/g\sim1.5\tevns$, where $g$ is the $SU(2)_L$ gauge
coupling.  While precision electroweak constraints disfavor strongly
coupled physics at this scale, various new models postpone unitarity
violation by introducing new weakly-coupled particles appearing at the
\tev scale.  This contribution reviews some of the recent experimental
developments in the planned searches for these particles at the ATLAS
and CMS detectors.  

\section{Technirho search at CMS}

Technicolor (TC) is one of the earliest models of strong symmetry breaking.  
Analogous to quantum chromodynamics, it introduces new massless
fermions (``technifermions'') whose chiral symmetry is spontaneously
broken by the formation of a condensate, which
is also responsible for the EWSB.  Three of the Goldstone
bosons (``technipions'') produced in the breaking of the chiral symmetry
provide the masses for the \Wpm ~and \Zz ~bosons.

The masses of the Standard Model (SM) fermions can be introduced by embedding
color, technicolor and flavor into a larger gauge group, whose breaking
gives rise to massive gauge bosons that mediate transitions between SM
fermions and technifermions.  For such ``extended TC'' interactions
not to lead to significant quark mixing that is inconsistent with limits
from flavor-changing neutral currents, the technicolor gauge coupling is
required to run very slowly (``walking TC'').  This requirement is
satisfied by having many technifermions, with lightest TC resonances
appearing below 1\tev\cite{TC}.  

The recent CMS study~\cite{cmsnote} examines one such technihadron,
color-singlet \rhotc, within the phenomenological framework of the
``technicolor straw man model''~\cite{StrawMan}.  The model assumes that
(i) the lowest-lying bound states of the lightest
technifermions can be considered in isolation, (ii) the isotriplet
($\Pi_{TC}^{\pm,0}$)
comprised of the lightest pseudo-scalar bound states are two-state
mixtures of the longitudinal $\Wpm_L$, $\Zz_L$ and mass-eigenstate
pseudo-Goldstone technipions.

The particular channel studied is
$\rhotc\to\Pi_{TC}\Pi_{TC}\to W+Z\to3\ell+\nu\,,\,
\ell=e\rm{\,or\,}\mu$, which has a final state cross-section 
($\sigma\times\BR$) of $1-370\,\fb$, as obtained from Pythia~\cite{pythia}
for 14 different selections of $(m(\rhotc),m(\pitc))$ within the range
$100\le m(\pitc)\le m(\rhotc)\le 600\gevns$.  The $\Pi_{TC}$ -- $\W_L$ mixing
is taken as $\sin\chi\sim 1/3$. The SM backgrounds from $WZ\to3\ell+\nu$,
$ZZ\to4\ell$, $\zbb\to2\ell+X$ and \tt ~pair production are considered,
all generated with Pythia except for the \zbb ~process, for which the
matrix elements are computed with CompHep~\cite{comphep}.  The detector
is simulated with the CMS fast simulation, FAMOS~\cite{famos}, validated
against Geant4-based simulation.

The \Z boson is reconstructed from a pair of same flavor and opposite
charge leptons, whose combined invariant mass is within 3 standard
deviations ($3\times2.6\gevns$) of the nominal \Z mass.  The \W~boson
is reconstructed from a third lepton and missing transverse energy in
the event by solving a quadratic equation
which uses the nominal mass of the \W~as a constraint.  The ambiguity
in the longitudinal direction of
the neutrino is resolved by choosing the solution with the lower
absolute longitudinal momentum ($|p_Z|$) for the neutrino.
All three leptons are required to have transverse momenta above certain
minima ($p_{T}^{\ell\,(1,2,3)}>(30,10,10)\gevns$) and to be isolated in the
detector -- a requirement which particularly aims to reduce the \zbb
~and \tt ~backgrounds.

As a final step to improve the signal to background ratio, kinematic
selection criteria are applied on the reconstructed vector boson
candidates: $p_{T}^{(W,Z)}>30\gev$ and $|\eta^W-\eta^Z|<1.2$.  The
latter requirement on the pseudorapidity difference is most useful
to reduce the background from the SM $ZW$ production, and has been
tested not to depend significantly on $m(\rhotc)$.

\vspace{-1.0pc}
\begin{figure}[h]
\begin{minipage}{16pc}
\begin{center}
\includegraphics[width=14pc,height=12pc]{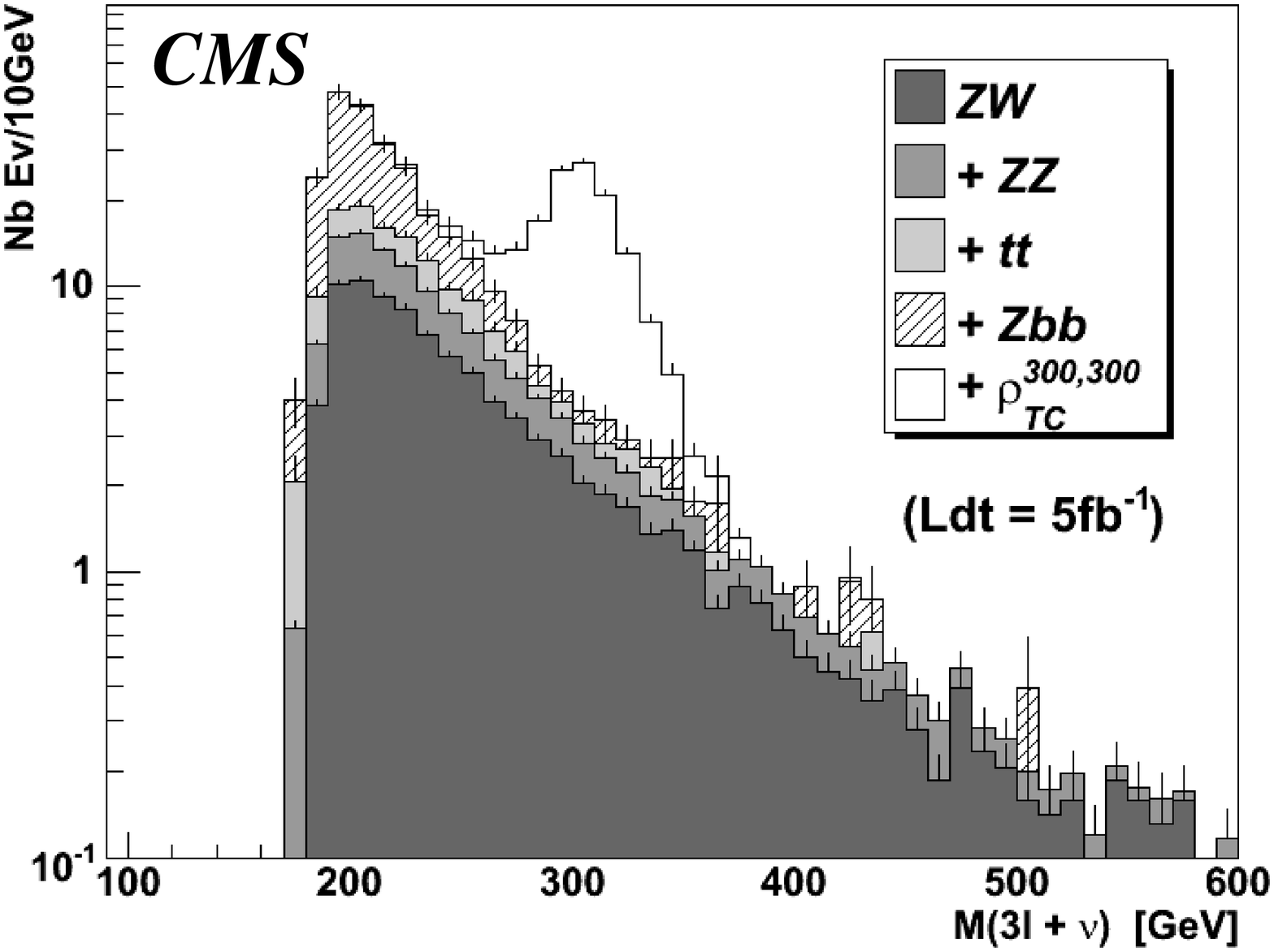}
\end{center}\vspace{-1.7pc}
\caption{\label{mtechnirho}\rhotc\,invariant mass distribution for 5\invfb of data.}
\end{minipage}\hspace{2pc}%
\begin{minipage}{18pc}
\begin{center}
\includegraphics[width=14pc,height=12pc]{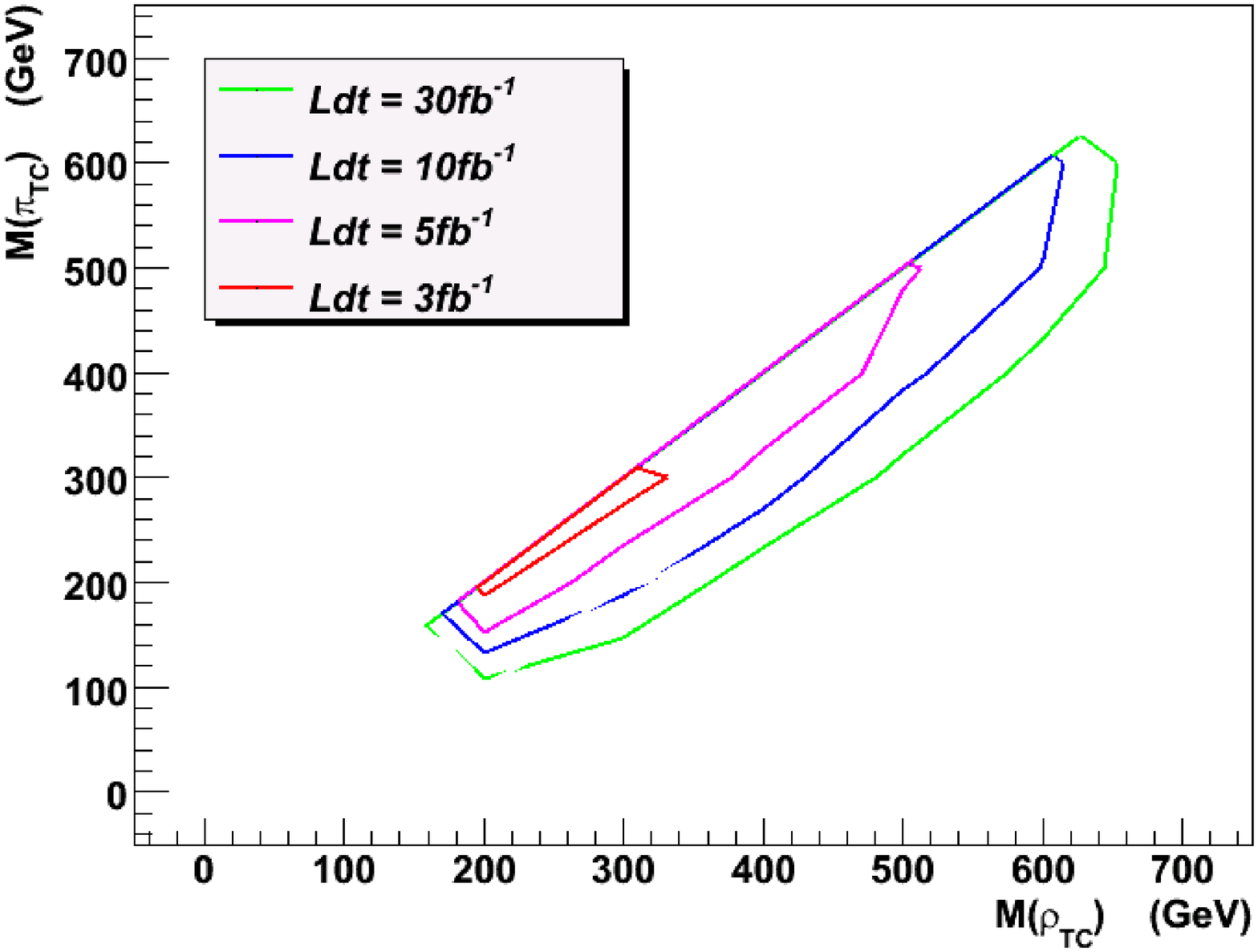}
\end{center}\vspace{-1.7pc}
\caption{\label{sigcontours}Contours of \rhotc\,discovery at $5\sigma$ significance, neglecting the systematic errors.}
\end{minipage} 
\end{figure}

The invariant mass of the \rhotc ~candidate reconstructed from the
selected vector boson candidates is shown in
Fig.~\ref{mtechnirho}, for the signal generated with
$(m(\rhotc),m(\pitc))=(300,300)\gevns$.  To determine the significance
of the signal above the background, a likelihood fit is performed with
the signal modelled with a Gaussian probability density function (pdf)
and the background with an exponential pdf.  A comparison of the
likelihoods for the signal-plus-background and background-only hypotheses
provides the significance estimate as
$S_{\Lh}=\sqrt{2\ln(\Lh_{S+B} / \Lh_{B})}$.  The expected
value of the estimate is obtained by multiple ``toy'' Monte-Carlo
experiments to be 7.7 for this particular choice of $(m(\rhotc),m(\pitc))$.

The $5\sigma$ discovery contours shown in Fig.~\ref{sigcontours} are obtained
by repeating the same procedure for different signals.  The sum of the
systematic uncertainties from various detector effects is about 11\%.  Their
effect on the contours is therefore small, with $5\sigma$ significance still
achievable with as low as 4\invfb of data for certain parts of the phase
space, after taking them into account.

\section{Vector boson scattering at ATLAS}

It is possible to study the EWSB in a model-independent way by removing
the Higgs field from the SM Lagrangian and introducing an extra field
with three degrees of freedom that will provide the masses of the
\Wpm~and \Zz~bosons.  This new field is chosen such that the 
Lagrangian is invariant under the full electroweak symmetry group. 
Prepared in analogy with low-energy QCD, the resulting effective theory
is called the Electroweak Chiral Lagrangian (EWChL)~\cite{kilian}.

The recent studies~\cite{theses}
~by the ATLAS Collaboration implement
the EWChL framework in the Pythia program to generate vector boson
scattering (VBS)
events.  When the $\Wpm/\Zz$ are on-shell, their quasi-elastic scattering
amplitude diverges at the lowest order unless there is some Higgs-like
mechanism.  Therefore VBS provides an excellent window to EWSB.  Imposing
CP-invariance and respecting precision electroweak constraints,
one finds that among all possible dimension-four-or-lower operators that
can be added to EWChL, only two contribute significantly to $\Wpm\Wpm$
~and $\Wpm\Zz$ channels.  The effective couplings for these terms,
\afour ~and \afive, determine if and where resonances should appear
in the mass spectrum, after the unitarization of the scattering amplitudes. 

The particular channel summarized in this contribution is
$q_1q_2\to q_3q_4\Wpm\Wpm$, with one of the final-state
\W~bosons decaying leptonically, and the other hadronically.  The bosons
are accompanied by two ``tag'' jets (from $q_3q_4$) at high
rapidity.  The considered SM backgrounds are from \W+jets and \tt, also
generated with Pythia.  ATLAS fast simulation program, ATLFast~\cite{atlfast},
is used for detector simulation.  The jets are identified using the $k_T$
algorithm~\cite{ktjet}.

Since the \W\W\,center-of-mass energy of interest is ${\cal O}(1\tevns)$, the
bosons are produced at high transverse momenta (\pt).  The leptonically
decaying \W~is reconstructed the same way as described
in the previous section, while the hadronically decaying \W~is identified
as the highest-\pt ~jet in the event.  The invariant mass of this jet and
the scale at which its constituents are resolved into two subjets are
required to be consistent with the values expected from genuine
\W~decays~\cite{bcf}.  The latter quantity, $\pt^2y_{21}$, is expected to be
${\cal O}(m_W^2)$ for the signal events, and the criterion
$1.55<\log(\pt\sqrt{y_{21}})<2.0$ can reduce the \W+jets background by an
additional 40\% after a $2\sigma$-mass-window requirement has been applied.

Both \W~candidates are required to have $\pt>320\gevns$.  To reduce the
\tt~background, each candidate is combined with any other jet in the
event and events having combinations close to the top-quark mass are
rejected.  The tag jets are identified as the highest-\pt~jets forward
and backward of the \W~candidates, and required to have pseudorapidity
$|\eta|>2$ and energy $E>300\gevns$.  With all four final state
objects identified, the total \pt~of the ``hard scattering'' system is
expected to be close to zero, so events with
$\pt(\W\W+{\rm tag jets})>50\gev$ are rejected.  Finally, events are
rejected when they contain more than one jet with $\pt>20\gevns$ which lies
between the two {\W}s in pseudorapidity, since QCD radiation is suppressed
in the central region in the signal with respect to the background.

\vspace{-0.2pc}
\begin{figure}[h]
\hspace{0.2pc}
\includegraphics[width=23pc,height=9pc]{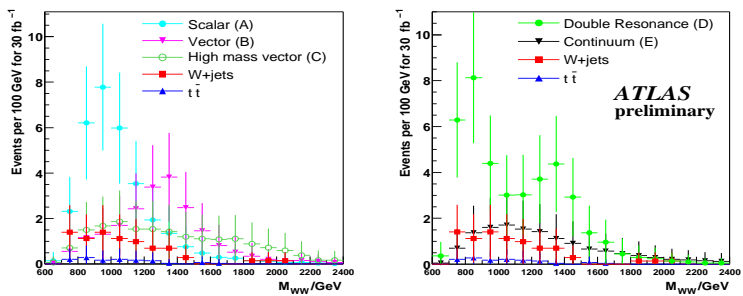}\hspace{1.2pc}%
\begin{minipage}[b]{12.2pc}\caption{\label{mww}Reconstructed \W\W~mass
distribution for five different signal scenarios (circles and downward
triangles) and the two backgrounds (squares and upward triangles) after
all cuts.}
\end{minipage}
\end{figure}
\vspace{-0.2pc}

Fig.~\ref{mww} shows the reconstructed \W\W~mass after all cuts for
five different signal scenarios.  In all cases, the signals are clearly
observable above the \tt~and \W+jets backgrounds for an integrated
luminosity of $30\invfb\!$.  Even the non-resonant (continuum) signal
with the lowest predicted cross-section yields an expected significance
of $S/\sqrt{B}\simeq4.7$.  Finally, the studies on $ZW$ scattering
indicate that significant signals can also be observed with 100\invfb 
in the $ZW\to\ell\nu qq$ channel and with 300\invfb in the
$ZW\to3\ell+\nu$ channel.

\ack
I would like to thank Sarah Allwood for her generous help in the
preparation of the slides.

\end{document}